\documentclass[a4paper,11pt]{article}

\usepackage{jheppub} 

\usepackage[T1]{fontenc} 

\title{A note about a new class of two-kinks}

\author[a]{T. S. Mendon\c ca}

\author[a,b,1]{H. P. de Oliveira \note{Corresponding author.}}

\affiliation[a]{
Departamento de F\'{\i}sica Te\'orica - Instituto de F\'{\i}sica
A. D. Tavares\\ Universidade do Estado do Rio de Janeiro, 
Rio de Janeiro, RJ, 20550-013, Brazil}

\affiliation[b]{Department of Physics and Astronomy \\
        University of Pittsburgh, Pittsburgh, PA 15260, USA}

\emailAdd{tiagobrouwer@msn.com}
\emailAdd{hp.deoliveira@pq.cnpq.br}

\abstract{
We present a model of two-kinks resulting from an explicit composition of two standards kinks of the $\phi^4$ model based on the procedure of Ref. \cite{uchiyama}.  The two-kinks have an additional parameter accounting for the separation of the standard kinks of $\phi^4$ model. We have shown that the two-kinks have two discrete internal modes besides the zeroth mode and the continuous spectrum. This new feature signalizes that the head-on collision a two-kinks/two-antikinks pair exhibits a rich and complex behavior due to the additional channel from which the energy of the system can be stored. We have exhibited the fractal structure associated with the main configurations after the collision. We have inferred the fractality as the imprint of the nonlinear exchange of energy into the two discrete internal modes.}

\begin{document} 
\maketitle
\flushbottom

\section{Introduction}

In a very interesting paper, Uchiyama \cite{uchiyama} had proposed an extended hadron model based on a modification of the sine-Gordon equation. The objective was to explain the quark confinement mechanism using two exact solutions: the first is for the modified sine-Gordon equation of the neutral scalar field, and the second is for the quark field equation in which the quark field interacts with the scalar field \cite{uchiyama}. The scalar field solution is obtained by an explicit composition of two kinks of the sine-Gordon model and for this reason Uchiyama named the new configuration as the two-kinks. 

Some years ago Bazeia et al. \cite{bazeia} introduced a new class of topological defects in systems described by real scalar fields in $(D,1)$ spacetime dimensions. By restricting to $D=1$ they have considered the family model with the potential,
\begin{equation}
V(\phi) = \frac{1}{2} \phi^2 (\phi^{-1/p}-\phi^{1/p})^2, \label{eq1}
\end{equation}
\noindent where the parameter $p$ is related to the way the scalar field self-interacts. This model can be understood as a generalization of the $\phi^4$ model, which is recovered for $p=1$. For $p$ even, the case $p=2$ is special and describes an unstable lumplike configuration \cite{bazeia}. For $p=4,6,..$ the potential (\ref{eq1}) requires that $\phi \geq 0$ or $\phi \leq 0$ under the change $\phi \rightarrow -\phi$. As pointed out by Refs. \cite{bazeia,bazeia_deformed} it is possible to construct topological defect in the form $\phi^{(+)}=\tanh^p(x/p)$ ($x \geq 0$) and $\phi^{(-)}=-\tanh^p(x/p)$ ($x \leq 0$). Moreover, non-topological lumplike defects can be envisaged according to Ref. \cite{new_lumplike}. It can be shown that all defects with $p$ even, topological or non-topological, are unstable and therefore of few interest.

The new static structures arising when $p=3,5..$ connect the minima $\phi=\pm 1$ passing through the symmetric minimum at $\phi=0$ are called two-kinks defects \cite{bazeia}. As shown in Fig. 1, the two-kinks defects seem to be composed of two standard kinks symmetrically separated by a distance proportional to $p$. The energy density profile reinforces this view. 

The static two-kinks solution associated to the potential (\ref{eq1}) is described by $\phi(x) = \pm \tanh^{p}\,(x/p)$. By implementing standard perturbation about the two-
kinks, $\phi(x,t)=\phi(x)+\mathrm{e}^{i \omega t} \chi(x)$, one can derive straightforwardly the following Schrodinger-like equation for the eigenmodes $\chi_n(x)$, 
\begin{equation}
\left[-\frac{d^2}{dx^2} + \left(\frac{d^2 V}{d \phi^2}\right)_K\right] \chi_n(x)=\omega_n^2\chi_n(x). \label{eq2}
\end{equation}
\noindent It happens that the effective potential diverges at the origin also shared by kinks of \cite{unusual_lumps}. In other words, the mass squared at $\phi=0$ diverges. This is not good for quantization since it cannot be used as a perturbation ground state. 

Two-kinks can be applied to the brane-world scenario \cite{brane_bazeia,chumbes,dutra}, and in connection with processes of their formation in perturbed sine-Gordon models \cite{monica}. Another possibility is to use two-kinks to describe the magnetic domain walls in constrained geometries \cite{dw_exps,dw_exps2}. Recently, we have studied numerically the head-on collision of two-kinks defects where we have noticed the appearance of unstable lump-like structures of non-topological nature as critical configurations \cite{deol_mendonca}. 

Although kink structures are ubiquitous in many branches of physics, particular interest relies on representing them as particles. In this context, kink-antikink configurations can mimic particle-antiparticle interactions and may provide insights towards fundamental concepts like crossing-symmetry in particle physics scattering \cite{weigel,cross}.

In this paper, we have extended this construction considering the two-kinks that arises from the union of two standard kinks of the $\phi^4$ model separated by a distance $R$. In Section 2, we have obtained the potential corresponding to the new two-kinks and discussed their stability. In Section 3 we have solved numerically the Schrodinger-like (\ref{eq2}) associated to the two-kinks to determine the internal modes. We have found that the two-kinks have two discrete internal modes besides the zeroth mode. In general, the eigenfrequencies $\omega$ depend on the separation parameter $R$. We have also shown that the two-kinks have a continuous spectrum of modes with eigenfrequencies $\omega \geq 2$ (cf. (\ref{eq3})). Motivated by the existence of the additional discrete mode, we have discussed the head-on collision of a two-kinks pair briefly in Section 3. We have exhibited a fractal structure with respect to the possible configurations after the collision. Finally, in Section 4 we summarize the main results and present a brief description of the further developments.

\begin{figure}[ht]
\begin{center}
\includegraphics[width=5.5cm,height=4cm]{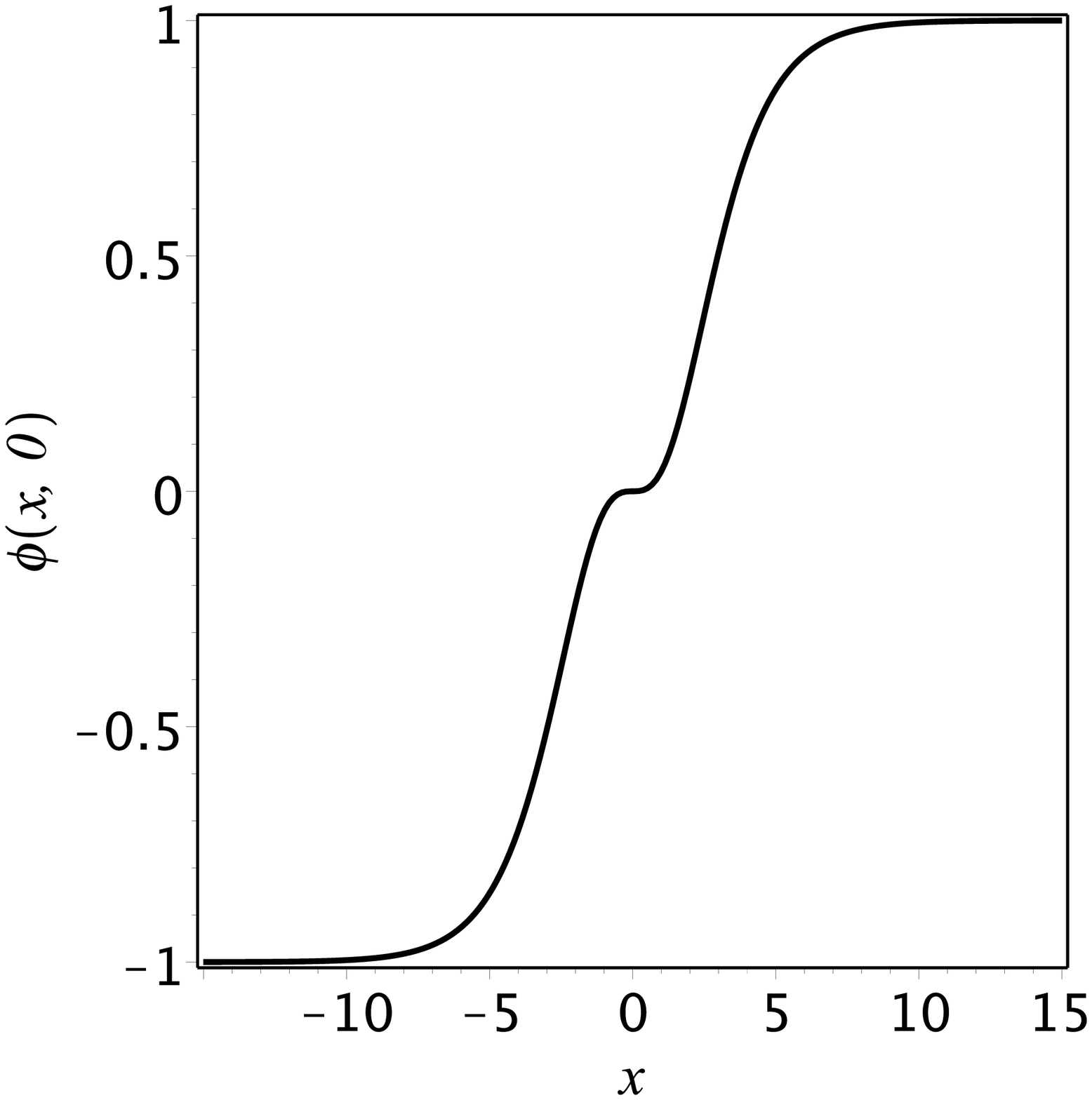}
\includegraphics[width=5.5cm,height=4cm]{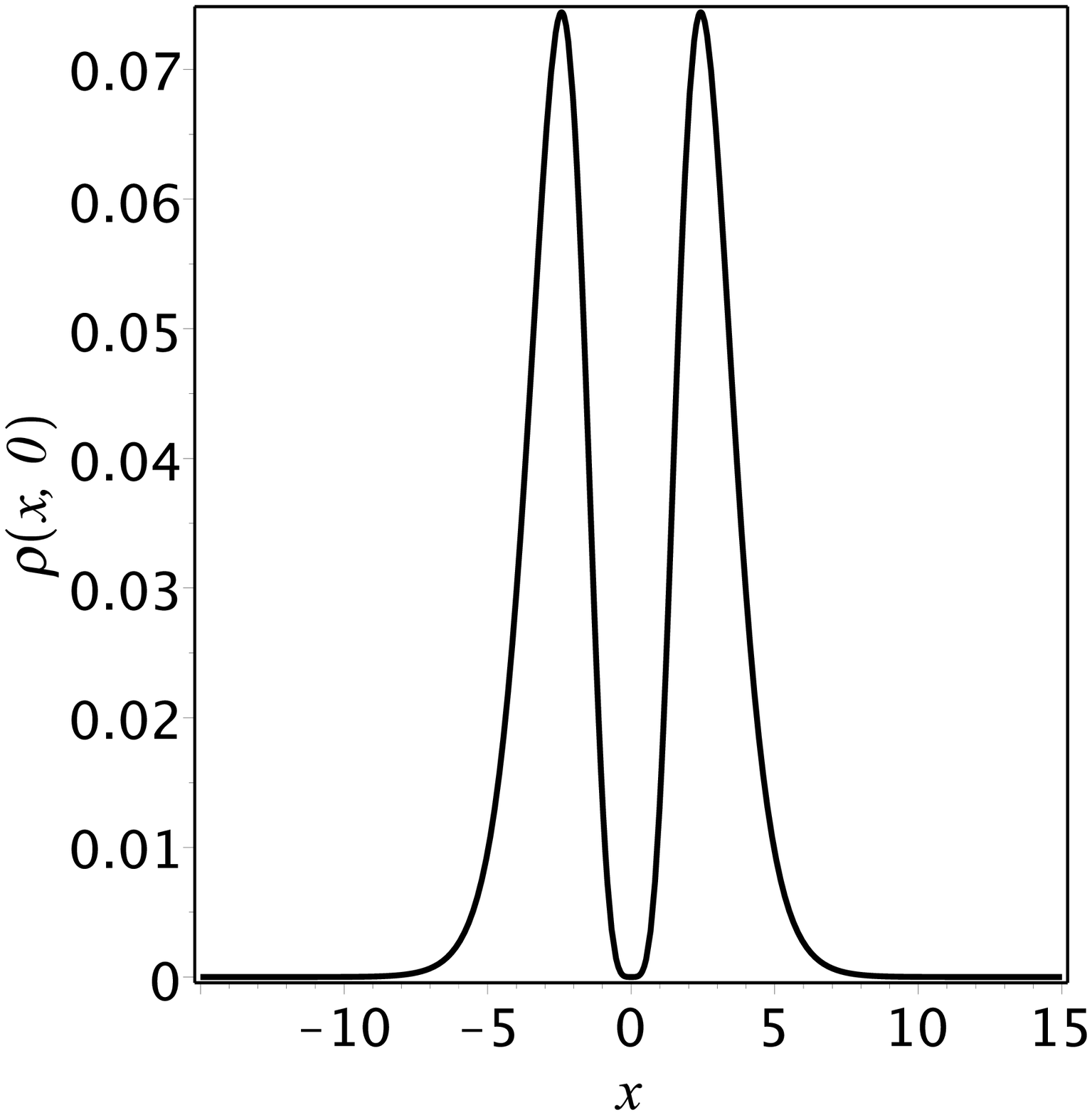} 
\end{center}
\caption{Profiles for the kink and its energy density for $p=3$ and $u=0.4$. The resulting structures are composed of two standard kinks symmetrically separated by a distance proportional to $p$ \cite{bazeia}.}
\end{figure}

\section{Uchyama's approach and the new two-kinks model}

Uchiyama had considered the sine-Gordon model whose potential is,
\begin{equation}
V(\phi) = \frac{m^4}{\lambda}\left[1-\cos\left(\frac{\sqrt{\lambda}}{m}\phi\right)\right],  \label{eq3}
\end{equation}
\noindent where for the sake of simplicity $m=\lambda=1$ so that the sine-Gordon kink solution reads as, 
\begin{eqnarray}
\phi_{\pm}(x)=4\tan^{-1}\left[\mathrm{e}^{\pm(x-x_0)}\right]. \label{eq4}
\end{eqnarray}    
\noindent Then, Uchiyama proposed a new kink solution by taking,
\begin{equation}
\phi(x) = \phi_{+}(x) - \phi_{-}(x). \label{eq5}
\end{equation}

\noindent The quantity $2R$ is the distance between the kinks, and $R$ is identified as the radius of the bag in the Uchiyama's model. After a straightforward calculation, it follows that, 
\begin{equation}
\tan\left[\frac{\phi(x)}{4}\right] = \frac{\tan(\phi_+/4)-\tan(\phi_-/4)}{1+\tan(\phi_+/4)\tan(\phi_-/4)}=\frac{\sinh(x)}{\cosh(2R)}. \label{eq6}
\end{equation}
\noindent The potential for the above solution obtained from the Klein-Gordon equation,
\begin{equation}
\frac{d^2\phi}{dx^2} = \sin \phi - (\cosh R)^{-2} \left[\sin \phi + 2 \sin\left(\frac{\phi}{2}\right)\right]. \label{eq7}
\end{equation}
\noindent We recover the sine-Gordon equation when the two-kinks are far away, i. e., for $R \rightarrow \infty$. The potential of the modified sine-Gordon equation becomes,
\begin{equation}
V(\phi) = 1 - \cos \phi + (\cosh R)^{-2} \left[3 + \cos \phi + 4 \cos\left(\frac{\phi}{2}\right)\right]. \label{eq8}
\end{equation}
\noindent The potential is invariant under the symmetry  $\phi \rightarrow \phi$ and has period $4\pi$. There is a high local minimum $8 (\cosh R)^{-2}$ at $\phi=0$. Uchiyama have chosen $R=5.06770956457570$ such that the potential has the values $0.0012687$ at $\phi=0$ and $2.0031718$ at $\phi=\pi$. Using value of $R$ the sine-Gordon and the above potentials almost coincide.

We can extend the above procedure to construct two-kinks using the standard kink solution of the $\phi^4$ model. Let us consider the single kink,
\begin{equation}
\phi_{\pm}(x)=\pm \tanh(x-x_0), \label{eq9}
\end{equation}
\noindent where $\pm$ indicate the kink/antikink solution. We propose the following two-kink solution:
\begin{eqnarray}
\phi(x) = \phi_{+}(x) - \phi_{-}(x)=\frac{\sinh(2x)}{\cosh(2R) + \cosh(2x)}, \label{eq10}
\end{eqnarray}   
\noindent where the parameter $R$ dictates the separation between the single kinks. If $R=0$ we recover the standard kink (\ref{eq9}). From the Klein-Gordon equation, the corresponding potential can be calculated straightforwardly resulting in the following expression,
\begin{equation}
V(\phi) = \frac{2(1-\phi^2)^2 \left[1+(\cosh^2 2R-1) \phi^2\right]}{\left[\cosh 2R +\sqrt{1+(\cosh^2 2R-1) \phi^2}\right]^2}.\label{eq11}
\end{equation}
\noindent It is instructive to expand the potential in power series of $R$,

\begin{equation}
V(\phi) = \frac{1}{2}\left({\phi}^{2}-1\right)^{2} + \left({\phi}^{2}-1\right)^3 R^2 + \mathcal{O}(R^4), \label{eq12}
\end{equation}

\noindent where we recover the potential of the $\phi^4$ model if $R=0$ as expected. The potential has two minima at $\phi=\pm 1$ and depending on the value of $R$ the origin can be a maximum, $0 \leq R < R_*=\cosh^{-1}(2)/2$, or a minimum if $R > R_*$. For the sake of illustration, we present in Fig. 2 the potential for several values of $R$ as well as the scalar field and the energy density profiles of the two-kinks.

We have noticed that the two-kinks (\ref{eq12}) have already been derived in a distinct context. Chumbes and Hott \cite{chumbes} have considered a model with two interacting scalar fields in brane world scenario with the introduction of a superpotential. They have constructed new models with one scalar field after establishing an orbit equation \cite{dutra} relating both fields. It turned out that one of the models of two-kinks coincides with we have obtained using the Uchyiama's approach.

\begin{figure}[ht]
\begin{center}
\includegraphics[width=5.5cm,height=4cm]{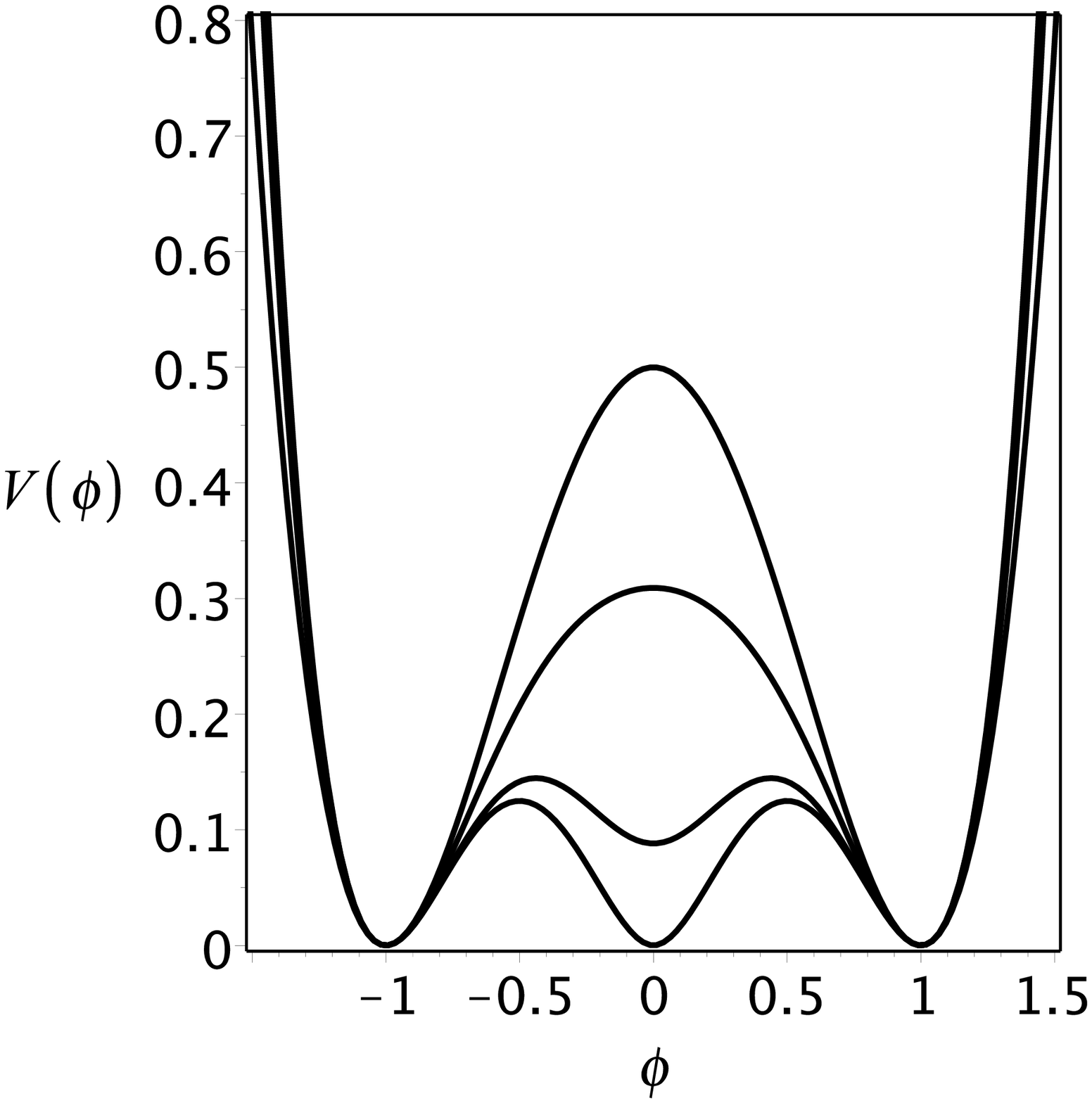}
\includegraphics[width=5.5cm,height=4cm]{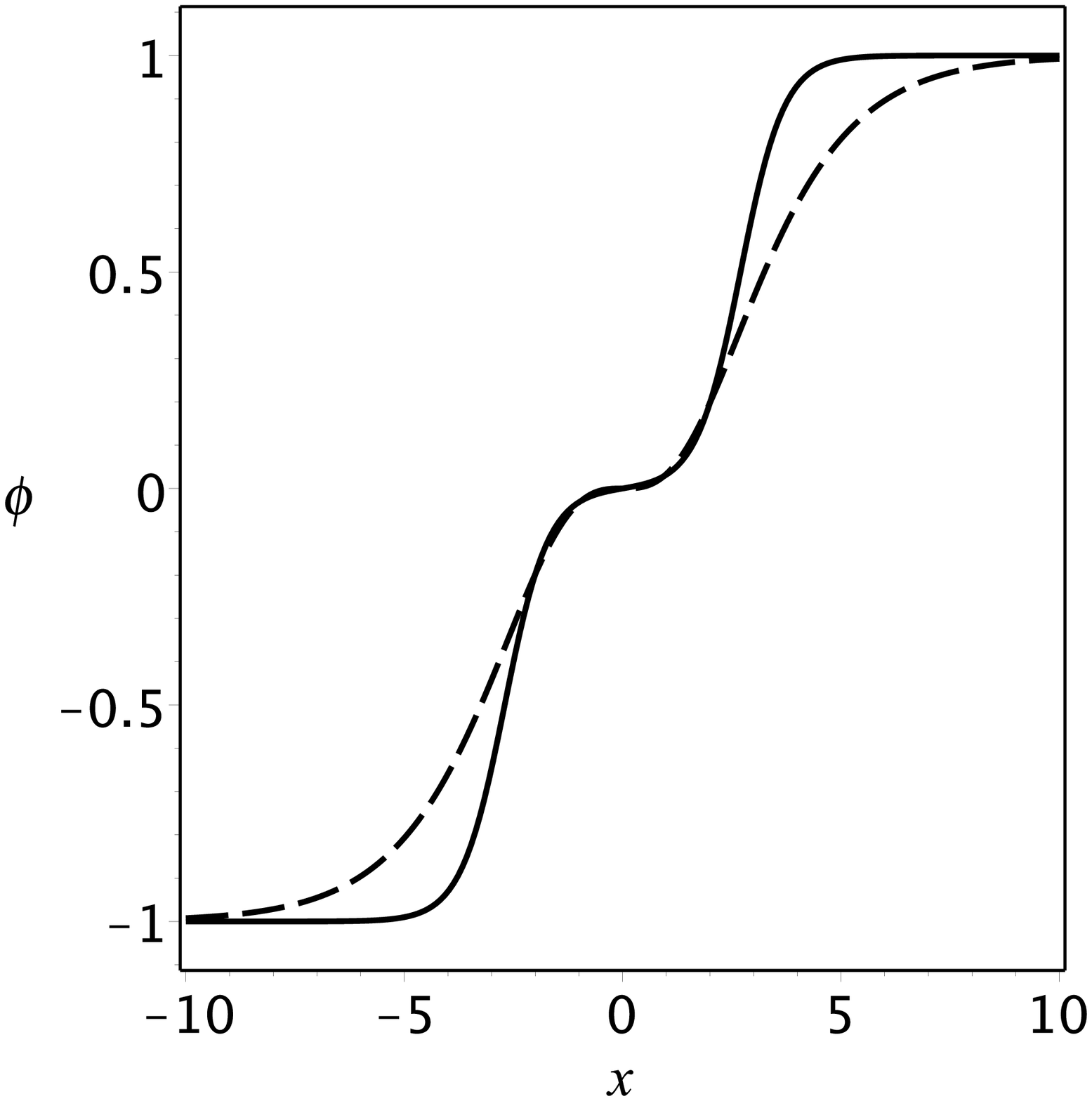} 
\includegraphics[width=5.5cm,height=4cm]{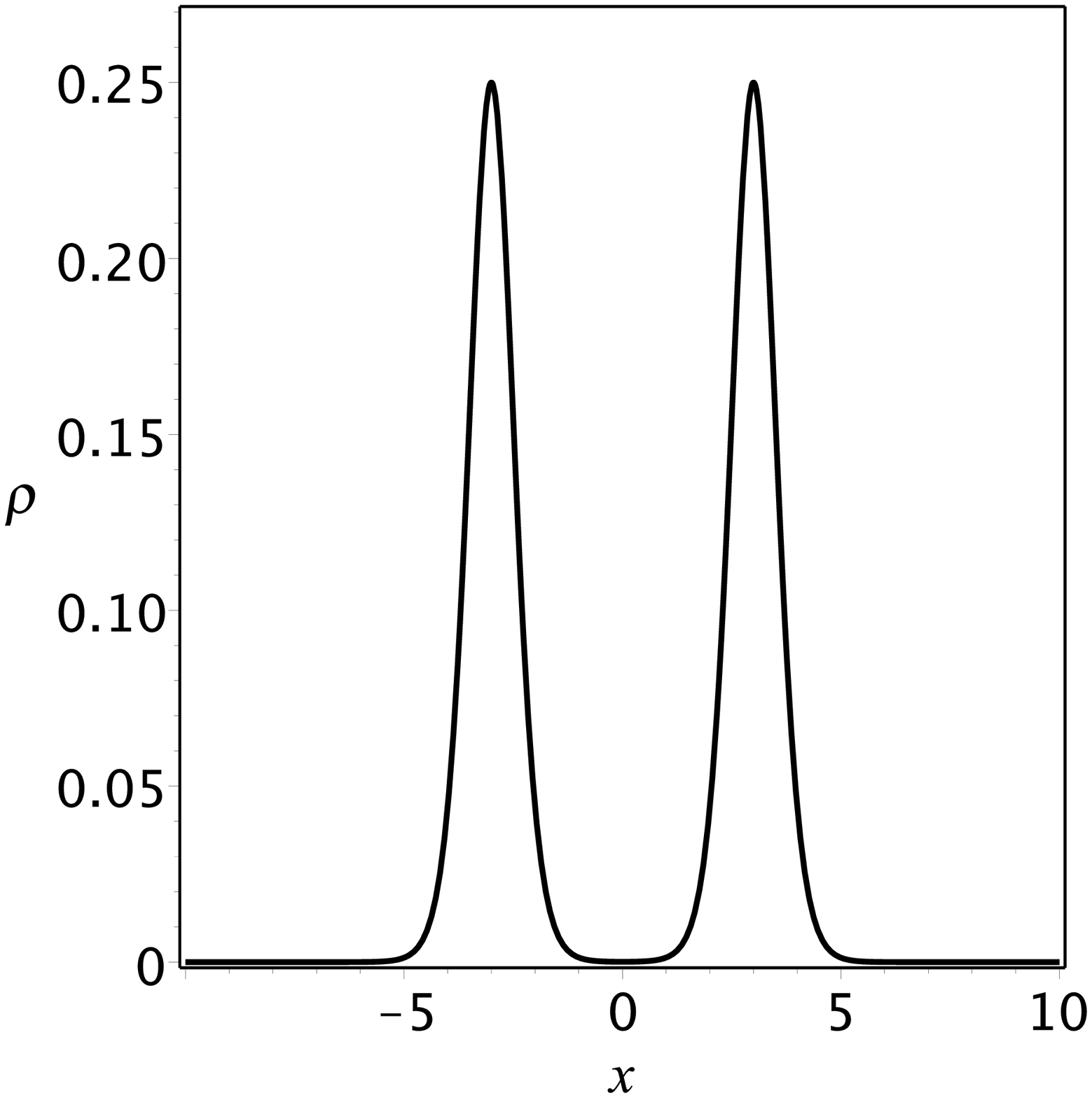} 
\end{center}
\caption{(a) Potential curves for $R=0,0.5,1.0$ and $3.0$, where for $R > 0.6584789484624083$ the origin turns to be a minimum. (b) Profiles of the two-kinks for $R=2.7$ and Bazeia's two-kinks \cite{bazeia} with $p=3$ (dashed line). (c) Typical energy density profile of a static two-kink with $R=2.5$.}
\end{figure}

The stability of the two-kinks (\ref{eq10}) can be inferred using the approach found in Ref. \cite{uchiyama}. In summary, it consists in perturbing the two-kink configuration by $\phi=\phi_0(x) + \epsilon \eta(x)$, where $\phi_0(x)$ is the two-kinks (\ref{eq12}), $\epsilon \ll 1$, and $\eta(x)$ is an arbitrary differentiable function satisfying the boundary conditions $\eta(\pm \infty)=0$. It can be shown that the total energy, 

\begin{eqnarray}
E(\epsilon) &=&  \int_{-\infty}^{\infty}\,\frac{1}{2}\left(\frac{d \phi}{dx}\right)^2+V(\phi) = \nonumber \\
&=& E_0 + \epsilon J_1 + \frac{1}{2}\epsilon^2 J_2 + \mathcal{O}(\epsilon^3),\label{eq13}
\end{eqnarray}

\noindent where the stability is fulfilled since $J_1=0$ and $J_2 > 0$.

\section{Internal modes of the two-kinks}

We can determine the internal modes of the two-kinks after solving the Schrodinger-like equation (\ref{eq2}). The zeroth mode or ground state $\omega=0$ has a known analytical solution regardless the kink model under consideration. One can show that, 

\begin{equation}
\chi_0(x) \propto \frac{d \phi_K(x)}{dx},\label{eq14}
\end{equation}

\noindent where $\phi_K(x)$ is given by Eq. (\ref{eq12}). In order to determine the other internal modes, we have to solve numerically Schrodinger-like equation (\ref{eq2}) to seek for regular solutions. 

At this point we recall that the $\phi^4$ model $(R=0)$ has one discrete mode with eigenfrequency $\omega = \sqrt{3}$. This discrete mode is the shape mode that plays a relevant role in explaining the bounces found in the kink-antikink collision. There is also a continuous spectrum with eigenfrequencies $\omega_k = \sqrt{k^2+4}$ known as the boson modes since they become plane waves asymptotically. 

We have integrated the Schrodinger-like equation using a fourth order Runge-Kutta algorithm to determine localized and well-behaved solutions. We have set the initial condition $(\chi(0) = 0,\chi^\prime(0)=0.2)$ and adjust the value of $\omega$ to obtain the desired eigensolutions. The novelty presented by the two-kinks is the existence of two discrete modes besides the zeroth mode. In addition, we have found a continuous spectrum with $\omega \geq 2$ no matter is the value of $R$. We have summarized the results in Fig. 3 with the values of $\omega$ in function of some values of $R$. For the sake of completeness, we have also displayed the results in Table 1. Notice that by increasing $R$, the highest discrete eigenfrequency tends to $\sqrt{3}$, whereas the smallest tends to zero. Therefore, for large $R$ the two-kinks have effectively only one discrete mode.  We have presented in Fig.4 the plots of the eigenfunctions of the two-kinks for $R=2$ together with shape mode of the $\phi^4$ model.

We have also considered the two-kinks associated to the potential (\ref{eq1}), where similarly to the standard kink of the $\phi^4$ model, there is only one internal discrete mode. For instance, if $p=3$ then $\omega \approx 0.55288762$; there is also a continuous spectrum for $\omega \geq 2$.

\begin{figure}[htb]
\begin{center}
\includegraphics[width=5.5cm,height=5.2cm]{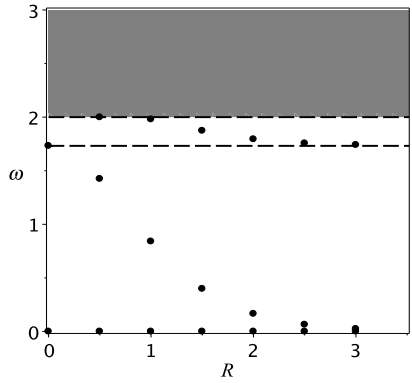}
\end{center}
\vspace{-0.5cm}
\caption{Eigenfrequencies $\omega_k$ corresponding to the internal modes $\chi_k(x)$ of the two-kink solution with respect to $R$. The dashed region $\omega>2$ denotes the continuous spectrum, and the dots indicate the discrete modes.}
\end{figure}

\begin{figure}[htb]
\begin{center}
\includegraphics[width=6.5cm,height=5.5cm]{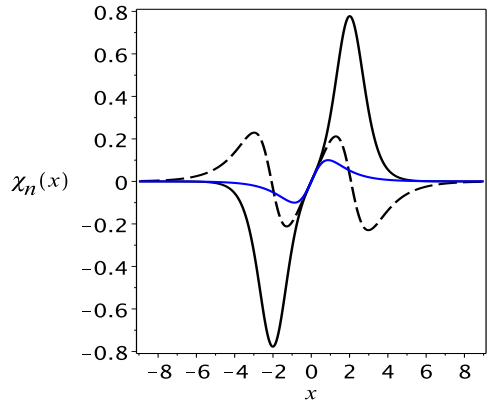}
\end{center}
\vspace{-0.5cm}
\caption{Discrete eigenfunctions of the two-kinks for $R=2$ corresponding to the frequencies $\omega = 0.165313536$ (continuous line) and $\omega = 1.79331783$ (dashed line). The blue line accounts for the shape mode of the $\phi^4$ where $\omega = \sqrt{3} \approx 1.73205081$.}
\end{figure}

\begin{table}
	\centering
		\begin{tabular}{c|c}
		\hline
		$R$ & $\omega$\\
		\hline
		\hline
		$0$ & $\sqrt{3} \approx 1.73205081$ \\
		\hline 
		$0.5 $& 1.42413098, $\approx 2.0$\\
	\hline
		1.0& 0.84029378, 1.97949332\\
	\hline
		1.5& 0.39826056, 1.87313903\\
	\hline
	2.0 & 0.16531354, 1.79331783\\
	\hline
	2.5 & 0.06398212, 1.75565933\\
	\hline
	3.0 & 0.02400534, 1.74077421\\
	\hline																		
		\end{tabular}
		\caption{Single precision values of the discrete eigenfrequencies corresponding to distinct $R$ shown in Fig. 3.}
\end{table}

\section{A glimpse on the two-kinks pair collision}

Let us present some preliminary result about the collision of a two-kinks pair. The initial configuration is,

\begin{equation}
\phi_0(0,x) = -1+\phi_K(0,x+x_0) - \phi_K(0,x-x_0), \label{eq15}
\end{equation}
\noindent where 

\begin{equation}
\phi_K(t,x) = \frac{\sinh 2 \xi(t,x)}{\cosh 2 \xi(t,x) + \cosh(2 R)},\label{eq16}
\end{equation}

\noindent with $\xi(t,x) = \frac{x-x_0-v t}{\sqrt{1-v^2}}$, represents the boosted two-kinks with velocity parameter $v$. We have integrated the Klein-Gordon equation with the above initial configuration using the spectral code of Ref. \cite{deol_mendonca}. We have set $x_0=15.0$ to guarantee that the two-kinks are sufficiently apart each other before the collision. In the numerical experiments, we have monitored the error rms in the energy conservation \cite{deol_mendonca} and verified that it is not superior to $0.01\%$. In order to achieve this error we have adopted $N=150$ in the spectral expansion of the scalar field, namely, $\phi(t,x)  = \sum_{j=0}^N\,a_j(t) \psi_j(x)$, where $a_j(t)$ are the unknown modes and $\psi_j(x)$ the basis functions (see Ref. \cite{deol_mendonca} for details).

There are two free parameters, the initial impact velocity and the separation parameter $R$. If we set large $R$, the second discrete internal mode is very close to zero, and the two-kinks have one discrete mode with $\omega \approx \sqrt{3}$. Nonetheless, the collision turns to be highly complex since the system behaves as a collision of four standard kinks of the $\phi^4$ model. Although we expect an intricate behavior, we will consider the collision of two-kinks with large $R$ elsewhere.

On the other hand, if $R$ is small, say $R \leq 3$, both discrete internal modes are effectively present. Several works have shown the role played by the shape mode in the collision of kinks in the $\phi^4$ model. Accordingly, this mode can absorb enough energy from the kink-antikink system preventing the kinks from moving apart after the collision and several bounces, or resonant configurations can occur \cite{moshir,fractal1,fractal2,aninos,goodman}. The bounces are a consequence of a competition of energy transfer to the shape and the translational mode. Therefore, the presence of at least one vibrational mode is crucial for the bouncing mechanism.

\begin{figure}[htb]
\begin{center}
\includegraphics[width=5.5cm,height=8.cm]{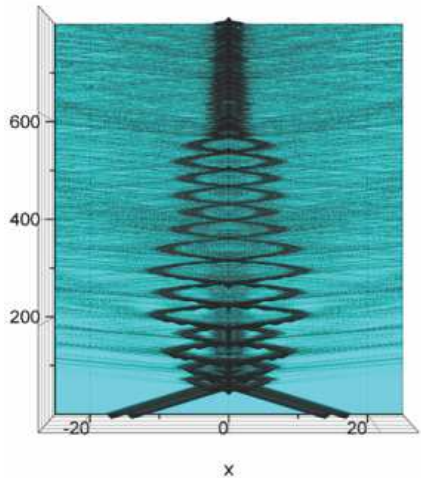}
\end{center}
\vspace{-0.5cm}
\caption{Three-dimensional plot of the energy density for the collision of a two-kinks pair for $u=0.23$, $R=1.5$ and initial separation $x_0=15.0$. There are fourteen irregular bounces before a bound oscillatory state forms at the origin.}
\end{figure}
 
We have fixed $R=1.5$ in all numerical experiments and evolved the system until $t_f = 600$. In this case, both discrete modes of the two-kinks might act as channels from which energy is transferred. As a consequence, we expect an intricate and complex behavior with the presence of multiple bounces and escape windows resulting from the nonlinear competition of the energy transfer to the discrete modes. In this vein, we expect some form of fractal pattern related to the configurations that emerge after the collision. However, we must recall that fractal patterns in the interaction of kinks of the $\phi^6$ model, although these kinks do not have a discrete internal mode \cite{phi6}. 

The numerical experiments have indicated that there exist a critical velocity, $v_c \approx 0.29$,  above which the collision is almost elastic, and both two-kinks escape to infinity after a brief interaction. Below the critical velocity, the collision is inelastic, therefore losing part of the scalar field. We have noticed the formation of a bound oscillating structure at the origin mixed with windows of escape. Moreover, multiple bounces may usually occur before the formation of an oscillon at the origin or when the two-kinks escape. For the sake of illustration, we have set $v=0.23$ to show the formation of multiple bounces before a bound solution at the origin is settled down. 

We can exhibit the fractal structure related to the main configurations that emerge from the two-kinks interaction for $v < v_c$. These configurations are: (i) the scalar field is trapped at the origin forming a bounded oscillating structure; (ii) the two-kinks escape to infinity; and (iii) two-kinks escape to infinity but leaving behind part of the scalar field trapped at the origin. In the dynamical system language, we call these states as basins of attraction. Then, in order to look up more carefully to the boundaries between these basis of attraction we have set the sequence of initial velocities $v=0.5,0.6,..,0.29,0.30$ and marked the corresponding basis of attractions (i), (ii) and (iii), respectively, by black, white and gray strips. These strips are centered about each velocity revealing a coarse distribution of the basis of attraction (cf. Fig. 6(a)). The next step consists in selecting a smaller interval of initial velocity, say from $v=0.13$ to $v=0.14$ that is indicated by a dashed rectangle shown in Fig. 6(a). We have simulated the two-kinks interaction with velocities inside this interval, $v=0.131,0.132,..,0.139$, and again placing the corresponding strips (cf. Fig. 6(b)). We have repeated further this procedure by taking smaller intervals of velocities, $0.130 \geq v \geq 0.131$ and $0.1301 \geq v \geq 0.1302$ and keeping track of the final configurations. We have presented the results in Figs. 6(c) and 6(d). The successive refinements establish quite clearly the non-smooth or fractal boundaries that separate the basins of attraction.

\begin{figure}[htb]
\begin{center}
\includegraphics[width=6.5cm,height=2.5cm]{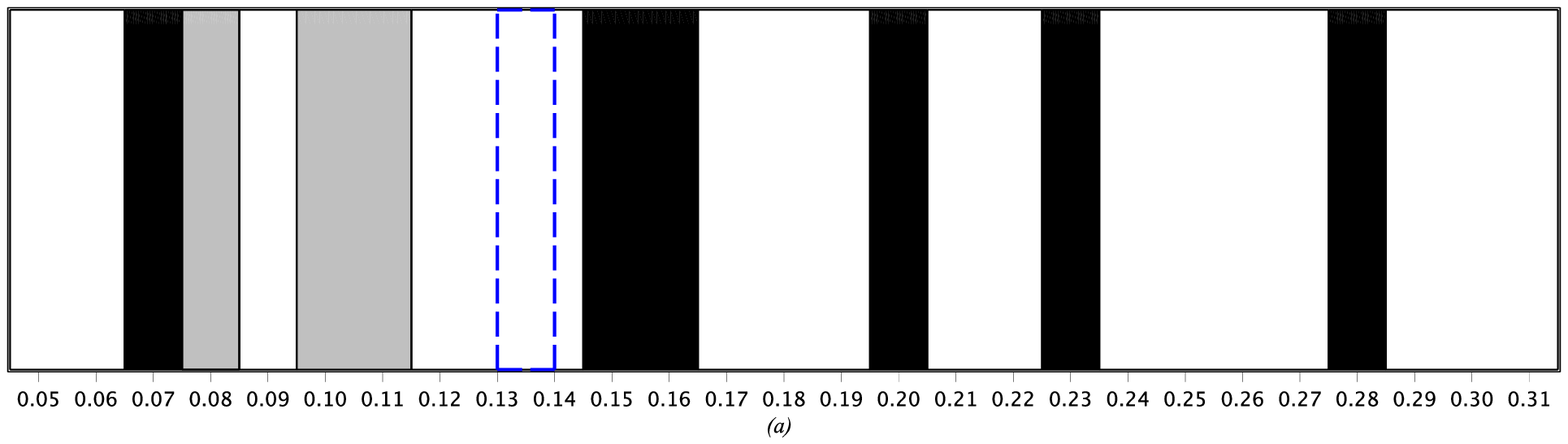}\\
\vspace{0.5cm}
\includegraphics[width=6.5cm,height=2.5cm]{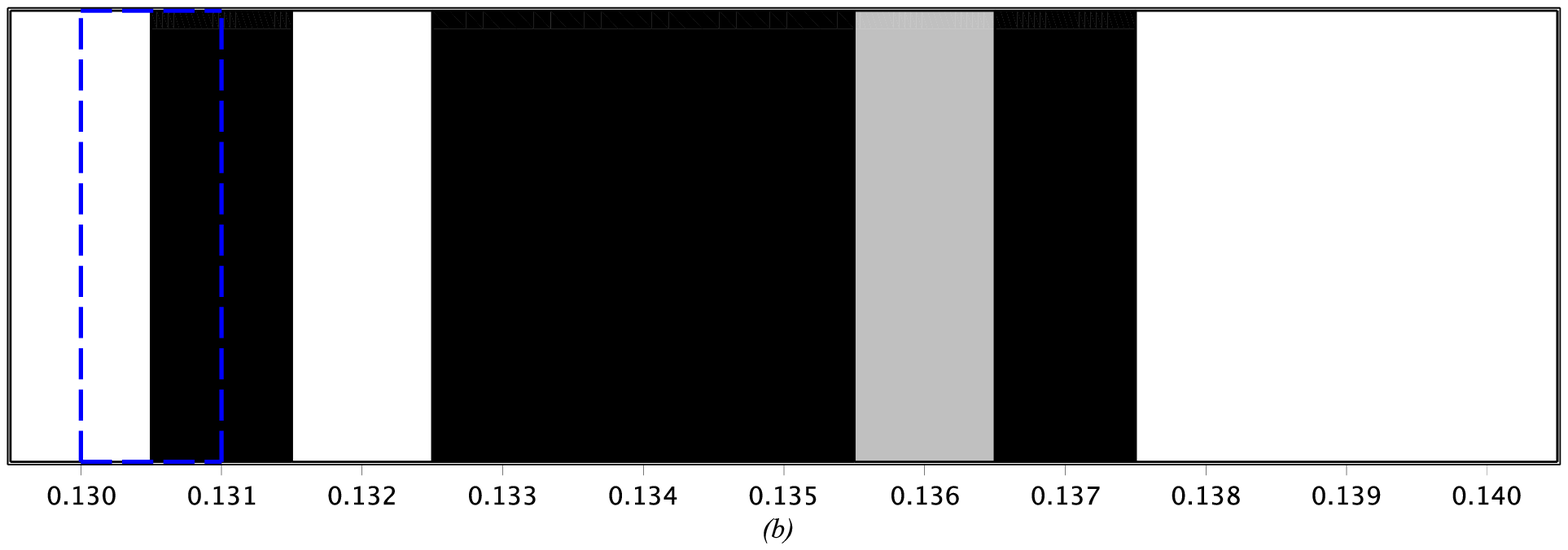}\\
\vspace{0.5cm}
\includegraphics[width=6.5cm,height=2.5cm]{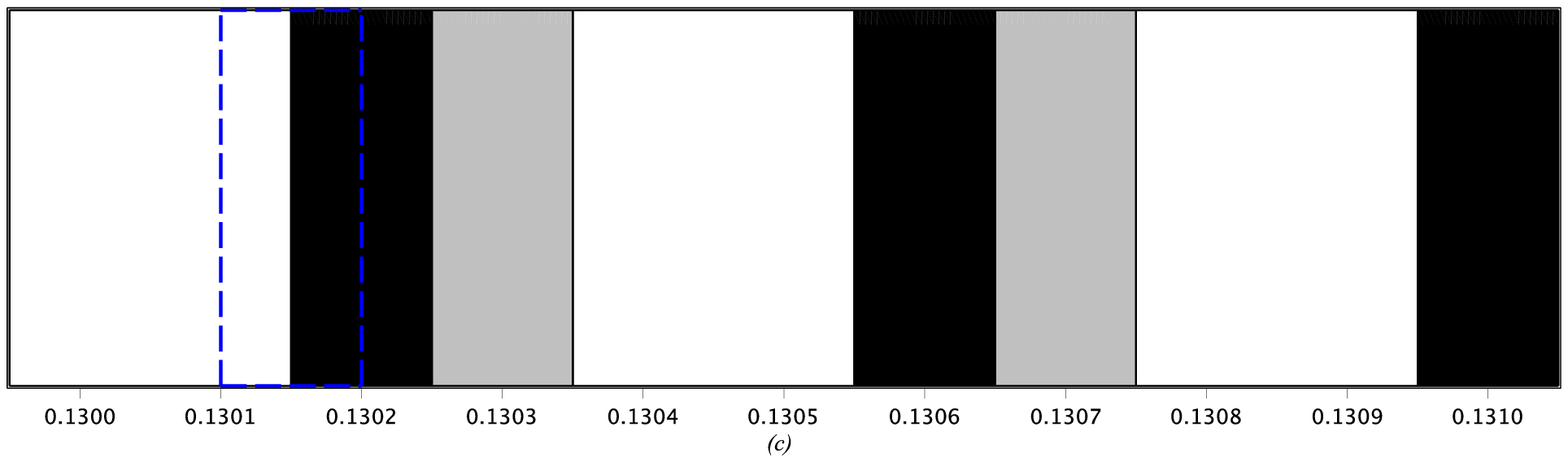}\\
\vspace{0.5cm}
\includegraphics[width=6.5cm,height=2.5cm]{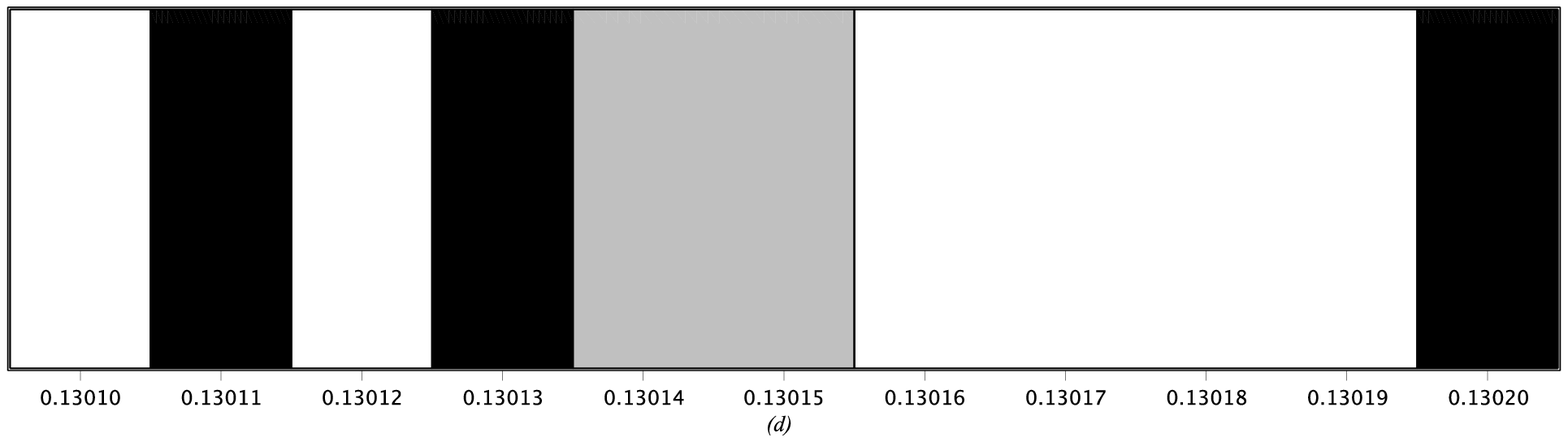}
\end{center}
\caption{Fractal structure related to the main configurations after the two-kinks interaction for $v < v_c$. The colors black, white and gray denote, respectively the following configurations: formation of an oscillon at the origin, escape of two-kinks, and escape of two-kinks together with an oscillon at the origin. Figure (a) provides a coarse distribution of these outcomes represented by strips centered at the velocities $v=0.05,0.06,..,0.38,0.29,0.31$. Figure (b) shows the structure within the interval $0.130 \geq v \geq 0.14$ indicated in (a) by a dashed rectangle. The successive refinements are present in Figs. (c), (d) showing that the boundaries between the main final configurations are fractal.}
\end{figure}

\section{Further perspectives}

In this paper, we have introduced a new model of two-kinks by extending the Uchyiama approach to the standard kinks of $\phi^4$ model. There is a parameter $R$ asserting the separation of the two standard kinks that forms the new structure. In this model, the effective potential of the Schrodinger-like equation for the internal modes is well behaved, contrary to the case of two-kinks with potential (\ref{eq1}). We have also briefly discussed the stability of the new two-kinks.

We have solved numerically the Schrodinger-like equation (\ref{eq2}) to determine the internal modes of the two-kinks. The novelty is the presence of two discrete modes besides the zeroth mode ($\omega=0$). We have also exhibited the influence of the parameter $R$ on the eigenfrequencies associated to the internal modes. The existence of a continuous spectrum for $\omega \geq 2$ is a common feature no matter value of the parameter $R$. We have also inspected numerically the Schrodinger-like equation for the two-kinks of the potential (\ref{eq1}) \cite{bazeia}, and we have found just one discrete internal.

We have studied the collision of two-kinks for $R=1.5$, where both  discrete modes are effectively present. We have found that if $v > v_c$ with $v_c \approx 0.29$, the collision is almost elastic and the two-kinks escape to infinity. On the other hand, for $v < v_c$ we have exhibited a fractal structure associated with the main configurations after the collision. Moreover, multiple bounces can intermediate any of the final configurations (cf. Fig.5). In general, possibly due to competition of  energy transfer from the system to both internal modes, we have observed multiple bounces and windows of escape.

This first study does not exhaust the complete understanding of the two-kinks collision. There are other models of two-kinks whose interaction could produce identical features we have observed such as several windows of escape. Further, we can gain analytical insight by using the collective coordinates approach in analogy to the studies of the $\phi^4$ model. In particular, the role played by the two discrete modes in the dynamics of the interaction of the two-kinks is a subject worth of investigation.  

\acknowledgments
The authors acknowledge the Brazilian agencies CNPq and CAPES for financial support.

\end{document}